\documentclass[preprint]{aastex6}
\usepackage{graphicx}
\usepackage{subfigure}
\newcommand{\suzaku}{{\sl Suzaku}}

\newcommand{\mkcflow}{{\sl mkcflow}}
\newcommand{\fea}{FeI-K$\alpha$}
\newcommand{\feb}{FeXXV-He$\alpha$}
\newcommand{\fec}{FeXXVI-Ly$\alpha$}
\newcommand{\subplot}[1]{\subfigure{\includegraphics[height=3in,width=2in,angle=270]{#1.ps}}}
\begin{document}
\title{An Empirical Correlation of $T_{\rm max}$ - $M_{\rm WD}$ of Dwarf Novae and The Average White Dwarf Mass in Cataclysmic Variables in the Galactic Bulge}
\author{Zhuo-li Yu}
\affil{School of Astronomy and Space Science and Key Laboratory of Modern Astronomy and Astrophysics, Nanjing University, Nanjing, P. R. China 210093}
\author{Xiao-jie Xu}
\email{xuxj@nju.edu.cn}
\affil{School of Astronomy and Space Science and Key Laboratory of Modern Astronomy and Astrophysics, Nanjing University, Nanjing, P. R. China 210093}
\author{Xiang-Dong Li}
\affil{School of Astronomy and Space Science and Key Laboratory of Modern Astronomy and Astrophysics, Nanjing University, Nanjing, P. R. China 210093}
\author{Tong Bao}
\affil{School of Astronomy and Space Science and Key Laboratory of Modern Astronomy and Astrophysics, Nanjing University, Nanjing, P. R. China 210093}
\author{Ying-xi Li}
\affil{School of Astronomy and Space Science and Key Laboratory of Modern Astronomy and Astrophysics, Nanjing University, Nanjing, P. R. China 210093}
\author{Yu-chen Xing}
\affil{School of Astronomy and Space Science and Key Laboratory of Modern Astronomy and Astrophysics, Nanjing University, Nanjing, P. R. China 210093}
\author{Yu-fu Shen}
\affil{School of Astronomy and Space Science and Key Laboratory of Modern Astronomy and Astrophysics, Nanjing University, Nanjing, P. R. China 210093}

\begin{abstract}
The mean white dwarf (WD) mass in the Galactic bulge cataclysmic variables (CVs) was measured by applying the the shock temperature-WD mass correlation of magnetic cataclysmic variables (mCVs) to the Galactic Bulge X-ray Emission (GBXE) spectra. However, the resulting mean WD mass is lower than that of the local CVs. This discrepancy could be explained by the dominating sources in the GBXE are non-mCVs instead of mCVs. In this work, we conduct an thorough investigation on the X-ray spectra of local DNe from \suzaku\ archives, and derive semi-empirical correlations between the shock temperature $T_{\rm max}$, the flux ratio of \fec\ to \feb\ lines, and WD mass for quiescent, non-magnetic CVs. By applying these correlations to the GBXE, we derive the average WD mass of CVs in the Galactic bulge to be $0.81\pm 0.07M_\odot$. This value is consistent with previous optical measurements of WD mass in local CVs.

\end{abstract}
\section{Introduction}
A cataclysmic Variables (CV) is a binary star where a white dwarf (WD) accretes matter from a main-sequence or a sub-giant star via Roche-lobe overflow. Sub groups of CVs includes magnetic CVs (mCVs) and non-magnetic ones based on the magnetic field strengths of the WDs \citep{Warner1995,Frank2002}. About 10-20\% CVs are mCVs, the others are non-mCVs, more specifically, dwarf novae (DNe). With X-ray luminosity $\sim~10^{30-34}$erg s$^{-1}$, CVs are important X-ray emitters in the Galaxy. In a mCV, for example, an intermediate polars (IPs), matters from the companion star are heated by a strong shock and emit X-ray before falling onto the WD surface along the WD magnetic field lines. In a non-magnetic CV, on the other hand, X-ray photons are mainly from a boundary layer near the WD surface, where accreted matter are heated by either a strong shock or a series of weak shocks\citep[e.g.,][]{Frank2002}. For both sub-classes of CVs, the X-ray spectra include a continuum from multi-temperature optical-thin thermal plasma and prominent emission lines (e.g., the H-like and He-like Fe lines near $7$ keV). The spectra could be well fitted by an absorbed cooling flow model (\mkcflow\ in Xspec) with an additional intrinsic absorption in some cases \citep{Mukai2003}. 

The WD in a CV could hardly retain its accreted matter \citep[e.g.,][]{Prialnik1995,Yaron2005,Liu2016}. As a result, the present WDs in CVs could reflect the initial properties of WDs when the CV was formed, thus could be used to constrain (binary) star evolution theories. Moreover, CVs are closely related to the possible progenitors of type Ia supernovae, as both of them involves accreting WDs. Traditionally the WD masses in CVs were measured in the  optical/UV band \citep[e.g.,][]{Friend1990,Mason2001,Zorotovic2011}, and the results suggested an average WD mass of $0.83\pm0.23M_\odot$ in local CVs \citep[e.g.,][]{Zorotovic2011}. In recent years, X-ray spectral fitting method showed its power as it can directly measure the temperature of the shock-heated matter, and imply the WD mass by assuming the strong shock condition\citep[e.g.,][]{Frank2002} in mCVs:
\begin{equation}
  T_{\rm max} = \frac{3}{8} \frac{\mu m_H}{k} \frac{GM}{R}.
  \label{equ:ip}
\end{equation}

In the above equation , $T_{\rm max}$ is the shock temperature, $\mu$ is the mean molecular weight, $m_{\rm H}$ is the mass of H atom, $k$ is Boltzmann constant, $G$ is gravitational constant, $M$ and $R$ are mass and radius of the WD, respectively. For non-magnetic CVs, $T_{\rm max}$ is half of that value because half of the gravitational energy has been dissipated in the accretion disk. With additional $M_{\rm WD}$ - $R_{\rm WD}$ relation of WDs \citep{Nauenberg1972}:
\begin{equation}
  R_{\rm WD}=7.8\times10^8[(\frac{1.44M_\odot}{M_{\rm WD}})^{2/3}-(\frac{M_{\rm WD}}{1.44M_\odot})^{2/3}]^{1/2} cm,
\end{equation}
one can derive the WD mass through
\begin{equation}
  T_{\rm max} = \frac{3}{8} \frac{\mu m_H}{k} GM_{\rm WD} (7.8\times10^8[(\frac{1.44M_\odot}{M_{\rm WD}})^{2/3}-(\frac{M_{\rm WD}}{1.44M_\odot})^{2/3}]^{1/2})^{-1}.
  \label{equ:iptm}
\end{equation}

The above correlation have been tested in dozens of local mCVs by various works\citep[e.g.,][]{Suleimanov2005,Yuasa2010}. A recent one was done by \citet{Yuasa2010}, where the authors found a mean WD mass of $0.88\pm0.25M_\odot$ for local mCVs based on \suzaku\ observations.

On the other hand, the population properties of WDs in distant CVs remains unclear because neither optical or X-ray observations could provide a CV sample with distance over several hundred parsecs. The Galactic diffuse X-ray background, on the other hand, provides a natural sample for such a study. Discovered more than 30 years ago, the Galactic Ridge/Bulge X-ray Emission (GR/BXE) spans across the Galactic ridge and bulge in the energy range of 2-10 keV \citep[Here we follow the definition of][who define the GBXE as regions of $|l_{*}| < 0^{\circ}.6$, $1^{\circ} < |b_{*}| < 3^{\circ}$]{Nobukawa2016}. The 1Ms {\sl Chandra} observations on the 'Limiting Window' in the Galactic bulge suggested that the GBXE is dominated by discrete sources \citep{Revnivtsev2009}. The majority of which were proposed to be IPs, with a minor contribution from active binaries \citep[ABs, e.g.,][]{Revnivtsev2009,Hong2012}. Thus, the composite X-ray spectra of the GBXE contains information from all the CVs (and ABs) in that direction which are free from selection effects and form an unbiased CV sample. With the assumption of mCV dominating the GBXE, \citet{Yuasa2012} fitted the $E>$15keV spectra of GBXE with an IP model, and showed that the average WD mass in CVs is $0.66_{-0.07}^{+0.09}M_\odot$, which is close to $0.5-0.60M_\odot$ from INTEGRAL observations \citep{Krivonos2007, Turler2010}.

However, these results are only marginally consistent, if not inconsistent at all, with  optical measurements for local CVs \citep[$0.83\pm 0.23M_\odot $,][]{Zorotovic2011} or those of the Galactic center\citep[$\sim0.9M_\odot $,][]{Hailey2016}. In the Galactic center, the CVs in the crowded stellar environment may have experienced dynamical interactions which preferentially bring massive WDs into binaries, thus increased the mean WD mass in CVs. But the $0.2M_{\odot}$ discrepancy between local and the GBXE CVs is still confusing. The $T_{\rm max}$ - $M_{\rm WD}$ correlation have been tested in dozens of local mCVs and there is no reason why this correlation could not be applied to mCVs in the Galactic bulge. Recently, \citet{Xu2016} proposed that the GBXE in the Fe emission line range (6-8 keV) should be dominated by non-mCVs instead of mCVs. Similar conclusions were also drawn by \citet{Nobukawa2016} and \citet{Hailey2016} by comparing X-ray continuum of different sources and of the Galactic center. Thus, we should apply the $T_{\rm max}$ - $M_{\rm WD}$ correlation of non-mCVs to GBXE spectra to make a reliable measurement of average WD mass. Moreover, the accreted matter may exhibit a series of weak shocks in the boundary layer(BL) of non-mCVs, and the $T_{\rm max}$ - $M_{\rm WD}$ correlation may deviate significantly from the theoretical formula\citep{Frank2002}. In this work, we choose to firstly test the validity of the strong shock assumption based on X-ray spectroscopy of local non-mCVs, and build semi-empirical $T_{\rm max}$ - $M_{\rm WD}$ - $I_{\rm 7.0/I_{\rm 6.7}}$ correlations. Next we apply these correlations to the GBXE and constrain the average $M_{\rm WD}$ in the Galactic bulge. We use \suzaku\ archived sample of non-mCVs in this work because \suzaku\ can provide both a relatively self-consistent sample of non-mCVs and well studied data of the GBXE\citep[e.g.,][]{Yuasa2012,Nobukawa2016}. What's more, the spectra generally cover 0.3-40 keV, which is suitable for reliable $T_{\rm max}$ measurements.

The rest of the paper is organized as follows: In \S~2 we describe our sample and data analysis methods; We present our results in \S~3; In \S~4 we compare them with existing results and discuss the implications; And finally, in \S~5, we provide a short summary.

\section{Observations and data reduction}
The \suzaku\ X-ray Observatory contains two types of instruments: one is the X-ray Imaging Spectrometers \citep[XIS,][]{Koyama2007}, the other is the Hard X-ray Detector \citep[HXD,][]{Takahashi2007}. The XIS consists of four sensors: one is made of back-illuminated CCD(XIS-1), and the other three are made of front-illuminated CCDs(XIS-0, 2, 3). XIS-2 suffered a catastrophic damage on 2006 November 9 and no useful data was transferred since then.

We cross-correlate the Suzaku online archive \footnote{http://heasarc.gsfc.nasa.gov/W3Browse/suzaku/suzamaster.html} with \citet{Ritter2003}'s CV catalog to search for publicly available observations. A total number of 25 observations on 18 DNe were found. The basic informations on observed sources  are listed in Table \ref{tbl:dn}. This sample is certainly not a complete one. Never the less, it could provide us information of the shock temperatures of DNe. Previous works based on similar observations only included XIS data \citep[e.g.,][]{Byckling2010,Wada2017}, but the best fitted $T_{\rm max}$ is usually above 10 keV, thus HXD may be needed to put tighter constraints to $T_{\rm max}$. We then include both XIS and HXD data in this work. 

The data processing and spectral analysis are carried out using HEAsoft\footnote{https://heasarc.gsfc.nasa.gov/docs/software/lheasoft/}(version 6.17, \citealt{Arnaud1996}). The event files are reduced with the standard pipeline {\sl aepipeline} with the latest calibration files (XIS:20160607). For each XIS screening image, we extracted source events from a 200$\arcsec$ circle and background events from a 250-400$\arcsec$ annulus, excluding regions outside CCD or contaminating sources with the {\sl xselect} tools. The response files and ancillary files were generated by the {\sl xisrmfgen} and {\sl xissimarfgen}, respectively. For every source, we used the {\sl addascaspec} tool to combine the spectra and response files of XIS-0, XIS-2(if exists) and XIS-3. We regroup the spectra so that the signal-to-noise ratio of each spectrum exceeded three. For HXD data, background files were downloaded from \suzaku\ background FTP server\footnote{ftp://legacy.gsfc.nasa.gov/suzaku/data/background} and data were processed using the {\sl hxdpinxbpi} tool. Due to the low net counts of HXD for most sources, the HXD spectra were grouped so that there were 2-3 bins at least.

\section{Results}

\subsection{X-ray Spectroscopy of Quiescent DNe}
The X-ray spectra of quiescent DNe could be well fitted with the {\sl mkcflow} model in Xspec \citep{Mushotzky1988,Mukai2003}.  We choose combined energy ranges of 0.3-10.0keV for XIS and 12.0-50.0keV for HXD(if existed). We started by fitting the background-subtracted spectra with the model {\sl phabs}$\times${\sl (mkcflow+gaussian)}, or {\sl phabs}$\times${\sl pcfabs}$\times${\sl (mkcflow+gaussian)} if additional intrinsic absorption is needed\citep[see, e.g.,][]{Wada2017,Byckling2010}, where {\sl phabs}, {\sl pcfabs}, {\sl mkcflow} and {\sl gaussian} components accounts for the interstellar foreground absorption, the intrinsic absorption, the emission from the shock-heated plasma and the \fea\ fluorescent line, respectively. Examples of the best-fitting are plotted in Figure \ref{fig:spec}, and the results of $T_{\rm max}$ are listed in Table \ref{tbl:tm}. The last column of Table \ref{tbl:tm} shows the flux ratio of \fec\ and \feb\ lines from \citet{Xu2016}.

\subsection{The Empirical $M_{\rm wd}$-$T_{\rm max}-I_{7.0/6.7}$ Correlations for DNe}
To build a $M_{\rm wd}$-$T_{\rm max}$ correlation for DNe, we exclude several sources which do not belong to the DNe class. For example, GK Per was proposed to be an IP with DN-like outbursts \citep[see, e.g.,][]{Kim1992}, BF Eri was proposed to be an old nova exhibiting 'stunted' outbursts\citep{Neustroev2008}. In addition, only quiescent DNe are included in the following analysis because the $T_{\rm max}$ in other states would be different from that in quiescent state due to the altering of the structure of the accretion disk and BL in high accretion rates\citep{Wada2017}. What's more, the Obs-ID 109015010 of SS Cyg did not include HXD data, so we only use the results from Obs-ID 400006010 to represent $T_{\rm max}$ of SS Cyg. The data points of excluded sources are still plotted in the figures only for reference.

\begin{figure}[htbp]
  \centering
  \includegraphics[scale=0.2]{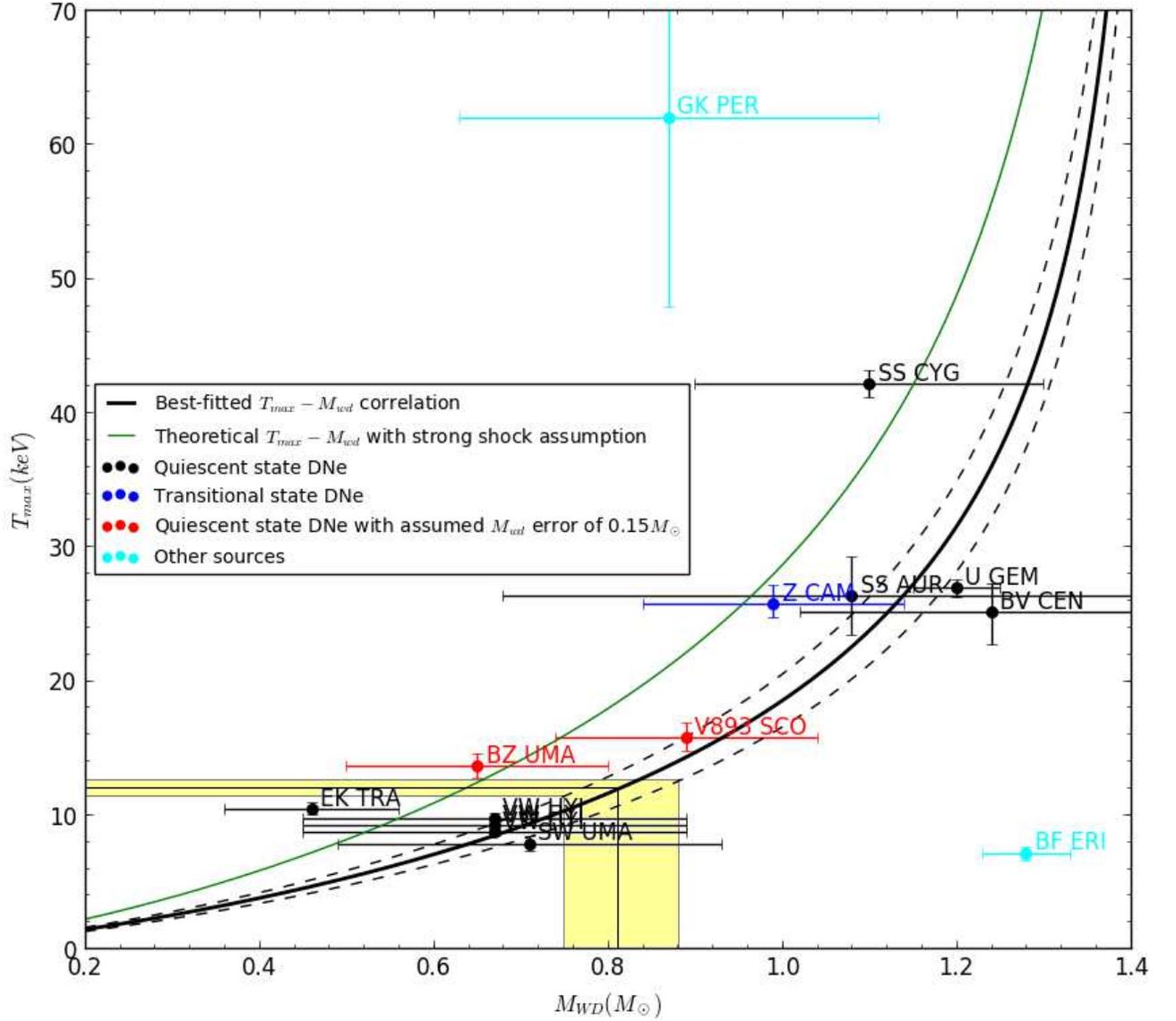}
  \caption{The $T_{\rm max}$ - $M_{\rm WD}$ correlation. Black points represent normal quiescent DNe. Red points are sources whose mass errors are assumed as 0.15$M_\odot$. Sources in transition state are plotted in blue points. Only black and red data points are used for fitting. The solid green curve shows the theoretical $T_{\rm max}$ - $M_{\rm WD}$ in strong shock assumption. The solid and dashed black curves show the best-fitted $T_{\rm max}$ - $M_{\rm WD}$ correlation and its error range.}
  \label{fig:tm}
\end{figure}

The masses of WDs in sampled DNe are adopted from various works, as listed in Table \ref{tbl:dn} (see references therein for details). We assign a typical 0.15$M_\odot$ uncertainty to WD masses of V893 Sco and BZ UMa because the mass errors were not mentioned in the references (this assumption would not affect our results, see \S4 for details). We further exclude CH UMa from our sample because its mass is higher than the Chandrasekhar limit. At last we have 11 available data points from 9 sources. In Figure \ref{fig:tm}, we plot $T_{\rm max}$ against the WD mass with a theoretical green curve representing the strong shock condition case. It is obvious that the green curve deviates from the data points. We then simply assume a parameter $\alpha$, so that $T_{\rm max}$ follows
\begin{equation}
  T_{\rm max} = \alpha \times \frac{3}{16} \frac{\mu m_H}{k} \frac{GM}{R}.
  \label{equ:fit}
\end{equation}

where $\mu$=0.6 is assumed. The best-fit results show an $\alpha=0.646\pm0.069$, with $\chi_{\nu}^2(DOF)=2.02(10)$ and $r^2$ value of 0.69. The fitted curve with error ranges is given as solid and dashed black curves in Figure \ref{fig:tm}.

\begin{figure}[htbp]
  \centering
  \includegraphics[scale=0.2]{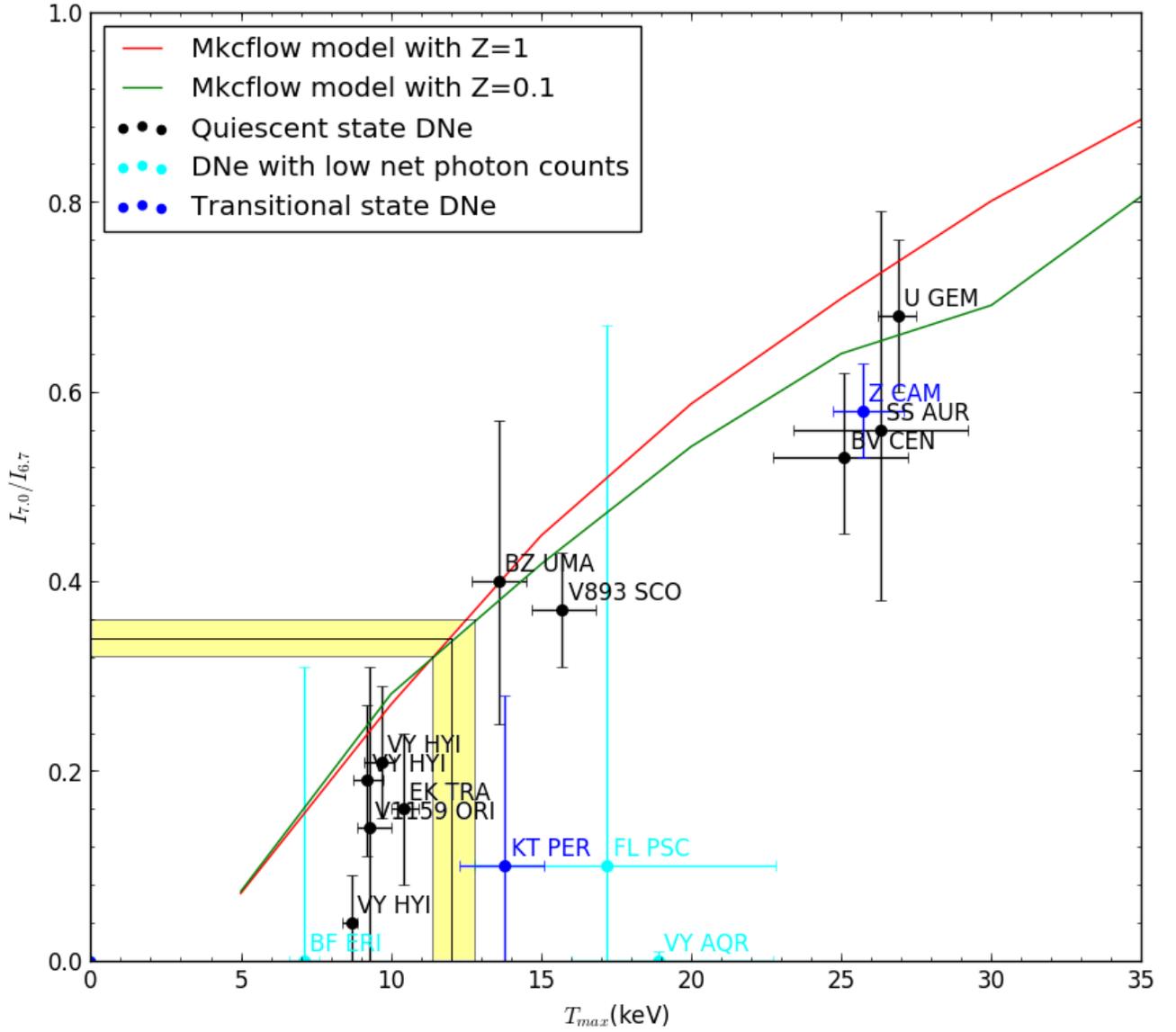}
  \caption{The $T_{\rm max}$ - $I_{7.0}/I_{6.7}$ correlation. The red and green curves show the theoretical correlation from $\sl mkcflow$ model for solar and $0.1$solar metallicity, respectively. The data points show in black, blue and cyan using $T_{\rm max}$ in this work and $I_{7.0}/I_{6.7}$ in \citet{Xu2016}. Points in black, blue and cyan are source in quiescent state, in transition state and those with relatively low net photon counts, respectively.}
  \label{fig:ratio}
\end{figure}

\begin{figure}[htbp]
  \centering
  \includegraphics[scale=0.2]{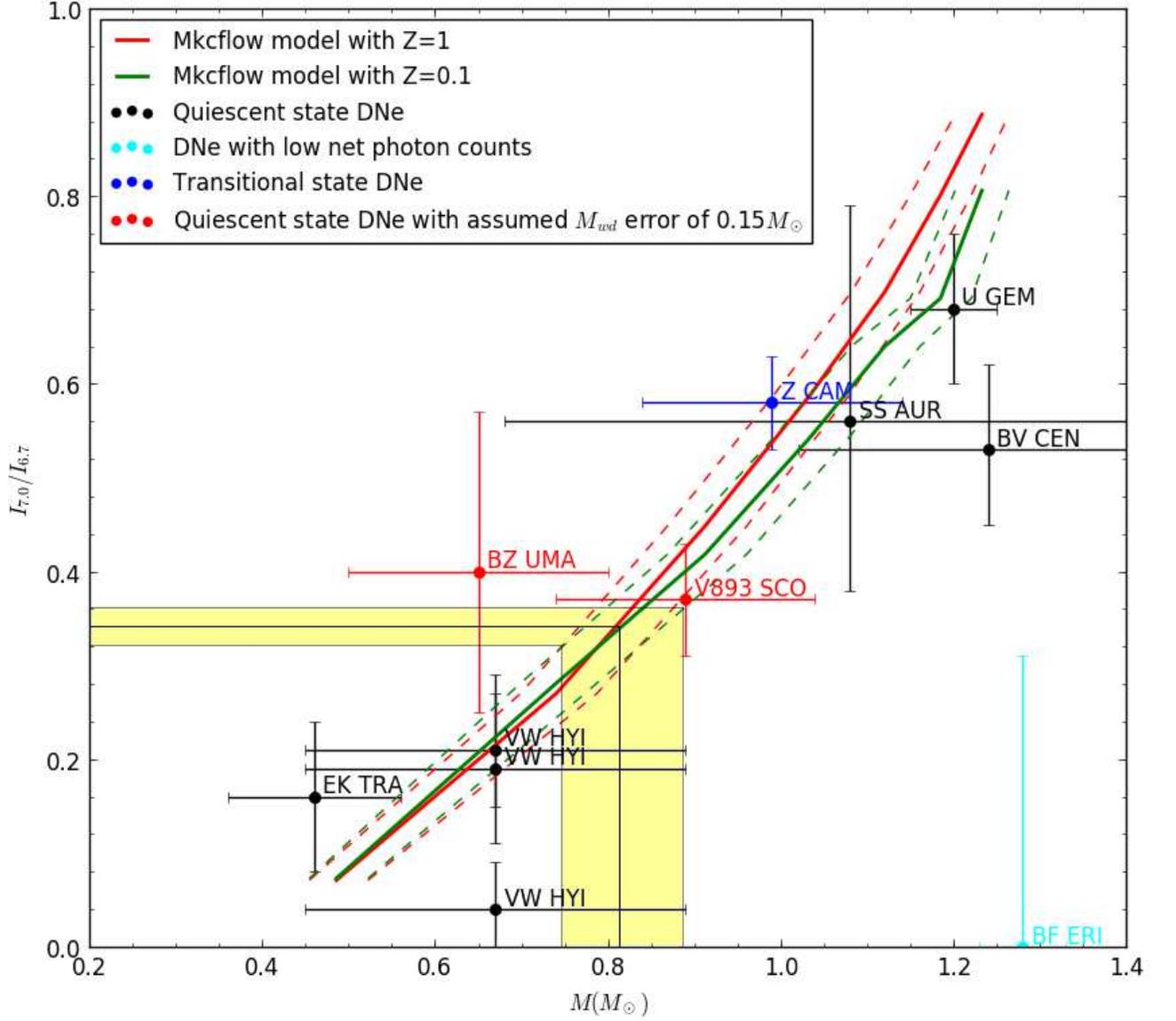}
  \caption{The $I_{7.0}/I_{6.7}$ - $M_{\rm WD}$ correlation. The curves show the semi-empirical correlation derived from equation  \ref{equ:fit} and $T_{\rm max}$ - $I_{7.0}/I_{6.7}$ theoretical correlation. The red and green curves correspond to abundance of 1 and 0.1 solar abundances, respectively. Normal DNe are plotted in black data points. Sources with assumed 0.15$M_\odot$ WD mass error are plotted in red points. The source in transition state are plotted as the blue point. Points in cyan show that sources have a relatively low net photon counts.}
  \label{fig:rm}
\end{figure}

As discussed in \citet{Xu2016}, $I_{7.0}/I_{6.7}$, the flux ratio of H-like Fe to He-like Fe lines, is a good indicator of $T_{\rm max}$ when $T_{\rm max}$ is below $\sim 50$keV. Thus $I_{7.0}/I_{6.7}$ is also correlated to the WD mass in CVs. In Figure \ref{fig:ratio} and Figure \ref{fig:rm}, we plot the measured $I_{7.0}/I_{6.7}$ against $T_{\rm max}$ and $M_{\rm WD}$. In addition, we simulate the the theoretical $I_{7.0}/I_{6.7}$ - $T_{\rm max}$ and $I_{7.0}/I_{6.7}$ - $M_{\rm WD}$ (where $M_{\rm WD}$ is converted to $T_{\rm max}$ using equation  \ref{equ:fit}, uncertainty of $\alpha$ has been considered) correlation curves for solar and $0.1$ solar metallicities from the {\sl mkcflow} model. The resulting curves are also plotted in the figures. The data points, in general, are consistent with the semi-empirical correlation curves, except three sources with low net counts (BF Eri, FL Psc and VY Aqr) and two sources in transition state (KT Per and Z Cam). We then conclude that DNe in quiescent state have $I_{7.0}/I_{6.7}$ - $T_{\rm max}$ and $I_{7.0}/I_{6.7}$ - $M_{\rm WD}$ correlations which are consistent with theoretical predictions. Moreover, the differences among different abundance is negligible when $T_{\rm max}$ is less than $\sim 15$ keV.

\section{Discussion}
\subsection{Comparison with Previous Measurements \& Bias Estimation}
Both \citet{Wada2017} and \citet{Byckling2010} have analyzed some of our sampled observations in their works.  We then compare our best fitting $T_{\rm max}$ with theirs in Table \ref{tbl:tm}. It is obvious that most of our results are consistent with theirs, which gives us confidence to make further investigations. The only exception is SS Cyg. Our $T_{\rm max}$ ($42.1\pm 1.0$keV) is consistent with that of \citet{Byckling2010} but is about $10$ keV lower than that of \citet{Wada2017}. We suspect the difference may be due to additional HXD data included in our analysis, because SS Cyg has a $T_{\rm max}$ much higher than 10 keV and should be better constrained by HXD data (see Figure \ref{fig:spec}). 

\citet{Byckling2010} have suggested that the correlation between $T_{\rm max}$ and $M_{\rm WD}$ in DNe is consistent with strong shock assumption, equivalent to $\alpha = 1$ in equation \ref{equ:fit}. However, our best-fitting results suggest $\alpha = 0.646\pm 0.069$. This discrepancy could be due to the different sampled DNe in their work. For example, BV Cen and U Gem are included in this work but not in theirs. These two sources are essential in our analysis because they provide important data points of relatively massive WDs, which could alter the fitting results significantly. What's more, \citet{Byckling2010} only plotted the  theoretical curve and did not provide a fitting, which may also affect their arguments. 

The potential bias in our fitting mostly come from the uncertainties brought by the optical/UV mass measurements and the limited sample size. As shown in Table \ref{tbl:dn}, the typical WD mass error is in the order of $\sim 0.1$-$0.2 M_{\odot}$. This error is equivalent to a $T_{\rm max}$ uncertainty of $\sim 8$-$20$ keV, which is much greater than the typical $T_{\rm max}$ error of $\sim 1$ keV in this work. On the other hand, only 9 out of 18 X-ray sampled DNe have mass measurements, which greatly limits our sample size and could add more uncertainties to our results. Thus, to test and improve the $T_{\rm max}$-$M_{\rm WD}$-$I_{\rm 7.0/6.7}$ correlation in the future, we need better constrained WD mass values of more CVs. What's more, there are multiple WD mass results for several sources, including SS Cyg and BZ UMa. For example, the WD mass in SS Cyg was measured to be  $0.81\pm 0.20M_{\odot}$, $1.09\pm 0.19M_{\odot}$ and $1.19\pm 0.02M_{\odot}$ by various authors \citep[e.g.,][]{Friend1990,Bitner2007}. We tested each of them in the fitting, and found the $T_{\rm max}$-$M_{\rm WD}$ correlation was altered by $\sim 10$\%, with $\alpha$ value from $0.58$ to $0.65$, respectively. We also tested the effects of excluding sources with assumed mass errors by removing data points of BZ UMa and V893 Sco, and performed fitting again. The new best fitted $\alpha$ value is $0.63\pm0.08$, which is within 5\% of the original value. These changes in $\alpha$ would alter the resulting mean $M_{\rm WD}$ (see \S4.2) in a $\sim~5\%$ level, so we will use $\alpha=0.646\pm 0.069$ to make further discussion. 

\subsection{Average Mass of WDs in DNe in the Galactic Bulge}
To imply the mean WD mass in the GBXE CVs, we need a reliable measurement of $T_{\rm max}$. However, the traditional method of deriving $T_{\rm max}$ by X-ray continuum fitting involves much complication in the GBXE case. GBXE continuum is a composition of multiple classes of sources \citep[e.g., mCVs, non-mCVs and ABs,][]{Revnivtsev2009}, and it is difficult to quantify the contribution of DNe alone. For example, in energy range above $10-15$ keV, the GBXE would be a mixture of both mCV and non-mCV emissions; in energy range below 2-4 keV, emission from ABs could be important\citep[e.g.,][]{Xu2016,Nobukawa2016}. So neither the hard X-ray band nor the broad band spectroscopy could give a reliable measurement of DNe $T_{\rm max}$. On the other hand, in the energy range containing He- and H-like Fe lines (6-8 keV), the GBXE spectra should be dominated by DNe \citep{Xu2016}. So we can make use of the Fe line flux ratio $I_{7.0}/I_{6.7}$ as an indicator of DNe $T_{\rm max}$, and imply $M_{\rm WD}$ directly. To do so, we adopt the GBXE Fe line ratio value of  $I_{7.0}/I_{6.7}=0.34\pm0.02$ from \citet{Nobukawa2016}. This value, according to the semi-empirical curves in Figure \ref{fig:ratio} and Figure \ref{fig:rm}, corresponds to a temperature of $T_{\rm max} = 12 \pm 0.6$keV and a mean WD mass of  $<M_{\rm WD}>=0.81\pm 0.07M_\odot$, respectively. This result is consistent with local value of $0.83\pm 0.23M_\odot$ by \citet{Zorotovic2011}.

The above measurement of $M_{WD}$ depends sensitively on precise $I_{7.0}/I_{6.7}$ value of DNe in GBXE. Recently, \citet{Nobukawa2016} conclude that, in the energy range of 5-10 keV, the flux percentage of mCVs, non-mCVs and ABs in GBXE are $\sim$3\%, $67\pm6$\% and $30\pm3$\%, respectively. Thus, the contribution of Fe lines from mCVs could be ignored. The contribution of Fe lines from ABs, on the other hand, should be considered. Although a quantitative investigation of \feb\ properties of ABs combined with their luminosity function is beyond the scope of this work, we could give a qualitative analysis as follows.  In general, ABs have relatively lower $T_{\rm max}$, thus emit more \feb\ photons and less \fec\ photons comparing to DNe, so the existence of ABs will reduce the $I_{7.0}/I_{6.7}$ value of GBXE. In another word, the real $I_{7.0}/I_{6.7}$ of DNe in GBXE should be higher than the measured GBXE value ($0.34\pm0.02$). As a result, our measured $<M_{\rm WD}>=0.81\pm 0.07M_\odot$ should be considered to be the lower limit of the average WD mass of DNe in the GBXE. This result is still consistent with \citet{Zorotovic2011}. 

Comparing to previous works of GBXE mean $M_{\rm WD}$, the mean WD mass of $0.81\pm 0.07M_\odot$ is higher than \citet{Yuasa2012}'s GBXE result of $\sim ~0.66M_{\odot}$. This discrepancy could be explained as follows. The \citet{Yuasa2012} work assumed a mCV origin of GBXE, and imply the mCV WD mass by $>15$keV GBXE X-ray continuum fitting. As discussed above, the GBXE in this energy range should include contribution of both mCVs and non-mCVs. Because non-mCVs usually have lower $T_{\rm max}$ than mCVs, \citet{Yuasa2012}'s fitting would result in a relatively lower $T_{\rm max}$, and therefore a $M_{\rm WD}$ lower than local mCVs\citep[$\sim 0.9M_{\odot},$][]{Yuasa2010}. Interestingly, directly applying the DNe $T_{\rm max}$ - $M_{\rm WD}$ correlation to the above hard X-ray continuum would instead overestimate the average WD mass to $\sim 1.15M_{\odot}$ (which is much higher than local value), because such a method would ignore the contribution of mCVs. In a word, the continuum fitting method can give robust measurements of $M_{\rm WD}$ of local CVs, but for the GBXE, the contribution of various classes of sources has to be quantified (e.g., a reliable luminosity function has to be built) before this method could be applied.

\section{Summary}
We have analyzed observations of DNe from \suzaku\ archives and presented an empirical formula of $T_{\rm max}$ - $M_{\rm WD}$ of DNe, which could help us estimate WD mass of DNe in quiescent state through X-ray observations. We also presents the $I_{7.0}/I_{6.7}$ - $T_{\rm max}$ and $I_{7.0}/I_{6.7}$ - $M_{\rm WD}$ correlation of DNe, and imply the average WD mass of DNe in the Galactic bulge to be  $<M_{\rm WD}>=0.81\pm 0.07M_\odot$. This result is consistent with measurements of local WDs in CVs.

\acknowledgements

This work is supported by National Science Foundation of China through grants NSFC-11303015 and NSFC-11133001.

\begin{table}[htbp]
  \caption{Sampled DNe Properties.}
  \centering
  \label{tbl:dn}
  \begin{tabular}{cccccc}
    \hline
    Source & Obs ID & Exposure time & Distance & $M_{\rm WD}$ & Reference\\
    & & (ks) & (pc) & ($M_\odot$) &\\
    \hline
    SS Cyg$_{1}$ & 109015010$^a$ & 43.2 & $117.1\pm6.2$ & $1.1\pm0.2$ & \citet{Ramsay2017}, \citet{Friend1990}\\
    SS Cyg$_{2}$& 400006010 & 39.5 &  $117.1\pm6.2$ & $1.1\pm0.2$ & \citet{Ramsay2017}, \citet{Friend1990}\\
    SS Cyg$_{3}$& 400007010$^b$ & 56.0 & $117.1\pm6.2$ & $1.1\pm0.2$ & \citet{Ramsay2017}, \citet{Friend1990}\\
    V893 Sco & 401041010 & 18.5 & $135_{-33}^{+65}$ & $0.89\pm0.15^{e}$ & \citet{Ozdonmez2015}, \citet{Mason2001}\\
    VY Aqr & 402043010 & 25.4 & $89_{-10}^{+13}$ & $-$& \citet{Ozdonmez2015}\\
    SW UMa & 402044010 & 16.9 & $159\pm22$ & $0.71\pm0.22$ & \citet{Gansicke2005}, \citet{Ritter2003}\\
    SS Aur & 402045010 & 19.5 & $167_{-9}^{+10}$ & $1.08\pm0.4$ & \citet{Ozdonmez2015}, \citet{Sion2008}\\
    BZ UMa & 402046010 & 29.7 & $204_{-37}^{+59}$ & $0.65\pm0.15^{e}$ & \citet{Ozdonmez2015}, \citet{Jurcevic1994}\\
    FL Psc & 403039010 & 33.3 & $160\pm40$ &$-$ & \citet{Patterson2011}\\
    KT Per & 403041010$^c$ & 29.2 & $145_{-21}^{+31}$ &$-$ & \citet{Ozdonmez2015}\\
    GK Per & 403081010 & 30.4 & $477_{-25}^{+28}$ & $-$& \citet{Harrison2013}\\
    Z Cam $_{1}$ & 404022010$^c$ & 37.7 & $219.3\pm19.5$ & $0.99\pm0.15$ & \citet{Ramsay2017}, \citet{Ritter2003}\\
   Z Cam$_{2}$ & 407016010$^b$ & 35.9 &$219.3\pm19.5$ & $0.99\pm0.15$ & \citet{Ramsay2017}, \citet{Ritter2003}\\
    VW Hyi$_{1}$ & 406009010$^b$ & 70.1 & $46\pm6$ & $0.67\pm0.22$ & \citet{Ozdonmez2015}, \citet{Hamilton2011}\\
   VW Hyi$_{2}$ & 406009020 & 16.2 & $46\pm6$ & $0.67\pm0.22$ & \citet{Ozdonmez2015}, \citet{Hamilton2011}\\
   VW Hyi$_{3}$ & 406009030 & 20.1 &$46\pm6$ & $0.67\pm0.22$ & \citet{Ozdonmez2015}, \citet{Hamilton2011}\\
   VW Hyi$_{4}$ & 406009040$^d$ & 16.8 &$46\pm6$ & $0.67\pm0.22$ & \citet{Ozdonmez2015}, \citet{Hamilton2011}\\
    U Gem$_{1}$ & 407034010 & 119.1 & $100\pm4$ & $1.2\pm0.05$ & \citet{Ozdonmez2015}, \citet{Ritter2003}\\
   U Gem$_{2}$  & 407035010$^b$ & 50.3 &$100\pm4$ & $1.2\pm0.05$ & \citet{Ozdonmez2015}, \citet{Ritter2003}\\
    CH UMa & 407043010 & 45.2 & $356\pm47$ & $1.95\pm0.3$ & \citet{Ozdonmez2015}, \citet{Friend1990}\\
    EK Tra & 407044010 & 77.8 & 180 & $0.46\pm0.1$ & \citet{Gansicke1997}, \citet{Mennickent1998}\\
    BF Eri & 407045010 & 32.8 & $596\pm79$ & $1.28\pm0.05$ & \citet{Ozdonmez2015}, \citet{Neustroev2008}\\
    BV Cen & 407047010 & 33.4 & $344.3\pm64.8$ & $1.24\pm0.22$ & \citet{Ramsay2017}, \citet{Watson2007}\\
    V1159 Ori & 408029010 & 200.5 & 299 &$-$ & \citet{ak2008}\\
    FS Aur & 408041010 & 62.2 & $246\pm32$ & $-$& \citet{Ozdonmez2015}\\
    \hline
  \end{tabular}
  
\scriptsize{The state is determined using the American Association of Variable Star Observers(AAVSO) International Database.\\
a. No HXD spectrum.\\
b. Outburst state.\\
c. Transitional state.\\
d. Very low HXD net counts, HXD data is not used.\\
e. $0.15M_{\odot}$ mass error assumed.}
\end{table}

\begin{table}[htbp]
  \centering
  \caption{Best-fit $T_{\rm max}$ \& Comparison with previous works.}
  \label{tbl:tm}
  \begin{tabular}{cccccccc}
    \hline
    Source & Id & $T_{\rm max}(keV)^a$ & $\chi_\nu^2(DOF)$ & Model & $T_{\rm max}(keV)^b$ & $T_{\rm max}(keV)^c$ & $I_{7.0}/I_{6.7}$ $^d$\\
    \hline
    SS Cyg & 400006010 & $42.1_{-1.0}^{+1.0}$  & 1.09(4513) & 1 & $52.5_{-0.7}^{+1.1}$ & $41.99_{-0.76}^{+1.20}$ &$-$\\
    V893 Sco & 401041010 & $15.7_{-1.0}^{+1.1}$ & 0.91(2767) & 2 & $19.2_{-0.6}^{+0.6}$ & $19.32_{-1.40}^{+1.29}$ & $0.37_{-0.06}^{+0.06}$\\
    VY Aqr & 402043010 & $18.9_{-2.9}^{+3.8}$ & 0.93(502) & 1 & $18.4_{-1.2}^{+2.3}$ & $16.47_{-2.22}^{+2.68}$ & $0_{-0}^{+0.01}$\\
    SW UMa & 402044010 & $7.8_{-0.5}^{+0.6}$ & 0.94(658) & 1 & $7.5_{-0.3}^{+0.4}$ & $8.33_{-0.99}^{+0.62}$ &$-$\\
    SS Aur & 402045010 & $26.3_{-2.9}^{+2.9}$ & 0.91(919) & 1 & $26.5_{-1.2}^{+2.1}$ & $23.47_{-3.02}^{+4.01}$ & $0.56_{-0.18}^{+0.23}$\\
    BZ UMa & 402046010 & $13.6_{-0.9}^{+0.9}$ & 1.00(1186) & 1 & $13.7_{-0.4}^{+0.6}$ & $13.71_{-0.81}^{+1.38}$ & $0.40_{-0.15}^{+0.17}$\\
    FL Psc & 403039010 & $17.2_{-4.4}^{+5.6}$ & 0.99(144) & 1 & $15.0_{-2.5}^{+1.7}$ & $14.43_{-2.69}^{+4.36}$ & $0.10_{-0.10}^{+0.57}$\\
    KT Per & 403041010 & $13.8_{-1.5}^{+1.3}$ & 0.92(1146) & 1 & $14.5_{-0.5}^{+0.4}$ &$-$ & $0.10_{-0.10}^{+0.18}$\\
    GK Per & 403081010 & $62.0_{-14.1}^{+14.5}$ & 0.94(1681) & 2 &$-$ & $-$& $0.94_{-0.45}^{+0.76}$\\
    Z Cam & 404022010 & $25.7_{-1.0}^{+1.4}$ & 1.10(3864) & 2 & $27.6_{-0.5}^{+0.3}$ & $25.76_{-2.39}^{+5.16}$ & $0.58_{-0.05}^{+0.05}$\\
    VW Hyi$_{1}$ & 406009020$^e$ & $8.7_{-0.3}^{+0.2}$ & 1.06(1518) & 1 & $9.3_{-0.1}^{+0.1}$ & $-$& $0.04_{-0.05}^{+0.05}$\\
   VW Hyi$_{2}$  & 406009030 & $9.7_{-0.6}^{+0.4}$ & 1.01(1524) & 1 & $10.0_{-0.1}^{+0.1}$ &$-$ & $0.21_{-0.06}^{+0.08}$\\
   VW Hyi$_{3}$  & 406009040 & $9.2_{-0.47}^{+0.55}$ & 0.98(1441) & 1 & $9.8_{-0.1}^{+0.1}$ & $-$& $0.19_{-0.08}^{+0.08}$\\
    U Gem & 407034010 & $26.9_{-0.7}^{+0.6}$ & 1.07(3387) & 1 & $26.2_{-0.6}^{+1.0}$ & $25.82_{-1.43}^{+1.98}$ & $0.68_{-0.08}^{+0.08}$\\
    CH UMa & 407043010 & $14.3_{-0.9}^{+0.7}$ & 0.99(1697) & 1 & $15.0_{-2.5}^{+1.7}$ &$-$ &\\
    EK Tra & 407044010 & $10.4_{-0.4}^{+0.5}$ & 1.07(2425) & 1 & $12.4_{-0.2}^{+0.1}$ &$-$ & $0.16_{-0.08}^{+0.08}$\\
    BF Eri & 407045010 & $7.1_{-0.5}^{+0.5}$ & 0.99(697) & 1 & $10.2_{-0.4}^{+0.9}$ &$-$ & $0_{-0}^{+0.31}$\\
    BV Cen & 407047010 & $25.1_{-2.4}^{+2.1}$ & 1.04(2866) & 2 & $27.5_{-0.7}^{+0.7}$ &$-$ & $0.53_{-0.08}^{+0.09}$\\
    V1159 Ori & 408029010 & $9.29_{-0.40}^{+0.74}$ & 1.01(1855) & 1 & $-$&$-$ & $0.37_{-0.06}^{+0.06}$\\
    FS Aur & 408041010 & $21.1_{-1.6}^{+2.0}$ & 0.96(1703) & 1 &$-$ & $-$&$-$\\
    \hline
  \end{tabular}
  
  \scriptsize{Model 1: \emph{phabs(mkcflow+gaussian)}, Model 2: \emph{phabs(pcfabs(mkcflow+gaussian))}\\
    Errors show 90 percent confidence level.\\
    a. This work, errors show 90 percent confidence level.\\
    b. Results of \citet{Wada2017}, errors are one standard deviation.\\
    c. Results of \citet{Byckling2010}, errors show 90 percent confidence level.\\
    d. $I_{7.0}/I_{6.7}$ refers to \citet{Xu2016}\\
    e. Energy band of $>$ 16keV for HXD is used.
 }
\end{table}

\begin{figure}[htbp]
  \centering
  \subplot{400006010}
  \subplot{401041010}
  \subplot{402044010}
  \subplot{402046010}
  \subplot{406009020}
  \subplot{407047010}
  \caption{Examples of broad-band DNe spectra with the best-fit models (see Table \ref{tbl:tm} for details). Obs-Ids and source names are as shown. Black data points stand for spectra from the combination of XIS-0, XIS-2(if existed) and XIS-3. Red and green data points represent spectra from XIS-1 and HXD, respectively. Lines are best-fitted models. Spectra are re-binned for plotting only.}
  \label{fig:spec}
\end{figure}

\end{document}